\documentstyle[amssymb,preprint,aps]{revtex}
\tightenlines
\begin{document}
\title{ A non-local, Lorentz-invariant, hidden-variable interpretation of relativistic quantum mechanics based on
particle trajectories}
\author{George Horton and Chris Dewdney}
\address{Division of Physics, University of Portsmouth. Portsmouth PO1 2DT. England}

\begin{abstract}
We demonstrate how to construct a lorentz-invariant, hidden-variable interpretation of relativistic quantum mechanics based on
particle trajectories. The covariant theory that we propose employs a multi-time formalism and 
a lorentz-invariant rule for the coordination of the space-time points on the individual 
particle trajectories. In this way we show that there is no contradiction between nonlocality
and lorentz invariance in quantum mechanics. The approach is illustrated for relativistic bosons, using a simple
model to discuss the individual non-locally correlated particle motion which ensues when the
wave function is entangled. A simple example of measurement is described. 

\end{abstract}
\maketitle

\section{Introduction}

In non-relativistic de Broglie-Bohm theory, when a many-particle wave
function is entangled the individual particle trajectories are non-locally
correlated. That is, the velocity of a given particle depends not just on
its own position but also on the simultaneous positions of all the other
particles at time $t$ as well. 
\begin{equation}
\overrightarrow{v}_{i}=V(\overrightarrow{x}_{1},\overrightarrow{x}_{2},....\overrightarrow{x}%
_{n},t)
\end{equation}
Thinking of the particles as instantaneously, non-locally correlated one
wonders what happens in relativistic theory where time-ordering is not
unique for space-like separated events. In which frame do you consider the
particles at the same time? There seems to be an apparent conflict between
non-locally correlated many-particle trajectories and the demands of
relativistic covariance. Hardy \cite{hardy1} has argued that any
hidden-variable approach to quantum theory will not satisfy Lorentz
invariance at the level of the hidden variables, although the statistical
predictions of relativistic quantum theory clearly will do so. Clearly the
issue can only be resolved properly within a covariant relativistic
hidden-variable theory for many particles. On the whole such relativistic
formulations as exist \cite{BohmandHiley}, \cite{Hollandb}(but see also
Duerr \cite{Duerr}, Goldstein \cite{Goldstein}, Holland\cite{Holland})
simply formulate the theory in a given frame.

As far as we know, although de Broglie formulated a theory for the Klein
Gordon equation \cite{deBroglie1}, he did not explicitly discuss this issue,
whereas Bohm suggested that his approach indicated the need for a preferred
inertial frame, with Lorentz invariance being satisfied \textit{only} at the
level of the statistical predictions of the theory. Valentini \cite
{Valentini} has suggested that the search for a Lorentz invariant
hidden-variable theory is in any case misguided as the fundamental symmetry
of the de Broglie-Bohm theory should be seen as Aristotelean. Our purpose in
this paper is to present an explicitly covariant de Broglie-Bohm theory, we
implement the model for the case of massive bosons, but the fundamental
principle can be used for fermion theories as well.

Several alternative ontologies have been proposed in the context of de
Broglie-Bohm theories. Bohm believed that bosons and fermions have a
fundamentally different character, in his approach, bosons are true fields
whereas fermions are true particles, both fields and particles are
well-defined and evolve in a continuous and causal manner according to
deterministic equations of motion. Valentini \cite{Valentinibook} has
suggested a theory in which both fermions and bosons are to be described by
well-defined fields. de Broglie preferred an ontology of particles for both
bosons and fermions, but for the boson case his prescription of the particle
trajectories had some serious pathologies. Elsewhere \cite{Dewdney96}, we
have discussed an outline of an approach which overcomes some of the
difficulties inherent in de Broglie's theory. More recently \cite{Horton
2000} we developed the approach in more detail demonstrating that it is
possible to develop a de Broglie-Bohm particle trajectory approach for the
Klein-Gordon equation which avoids the pathologies of de Broglie's original
theory. de Broglie's guidance condition for scalar bosons was based on the
charge current, whereas our approach uses the time-like flows of
stress-energy-momentum to define particle trajectories which follow the
flows.

\section{Stress-energy-momentum tensor for many particles}

Our Lorentz invariant description defines the flow of
stress-energy-momentum, and hence particle trajectories which follow the
flow, through the intrinsic natural four-vector provided by the matter field
itself through the eigenvalue equation 
\begin{equation}
T_{\nu }^{\mu }W^{\nu }=\lambda W^{\mu }  \label{eigenvalue}
\end{equation}
where $T_{\nu }^{\mu }$ is the stress-energy-momentum tensor, $\lambda $ the
eigenvalue and $W^{\mu }$ the eigenvector. If $\phi $ is a solution to the
Klein-Gordon equation then 
\begin{equation}
T_{\nu }^{\mu }=\frac{1}{2}\left[ \phi _{,\mu }^{\ast }\phi _{,\nu }-g_{\mu
\nu }\left( \phi ^{,\sigma }\phi _{,\sigma }-m^{2}\phi ^{\ast }\phi \right)
+\phi _{,\mu }\phi_{,\nu } ^{\ast }\right]
\end{equation}
Writing the solution $\phi $ as 
\begin{equation}
\phi =\exp [P+iS]  \label{eq:PSdecomposition}
\end{equation}
the stress-energy-momentum tensor, $T_{\nu }^{\mu },$ of the field $\phi $
is given by 
\begin{equation}
T_{\nu }^{\mu }=|\phi |^{2}[m^{2}-(P^{\alpha }P_{\alpha }+S^{\alpha
}S_{\alpha })]\delta _{\nu }^{\mu }+2|\phi |^{2}[(P^{\mu }P_{\nu }+S^{\mu
}S_{\nu })]  \label{eq:energy-mom}
\end{equation}
For an $n-$ particle system with no inter-particle interactions the
Lagrangian density can be taken to be 
\begin{equation}
L=\sum_{i}\frac{1}{2}\left( \phi ^{,\mu _{i}}\phi _{,\mu _{i}}^{\ast
}-m_{i}^{2}\phi \phi ^{\ast }\right)
\end{equation}
where $\phi =\phi \left( \overrightarrow{x}_{1},t_{1};....;\overrightarrow{x}%
_{n},t_{n}\right) $ and 
\begin{equation}
\phi _{,\mu _{i}}=\frac{\partial \phi }{\partial x_{i}^{\mu }}
\end{equation}
It then follows by the standard variational procedures that the field
equations are 
\begin{equation}
\left( \square _{i}^{2}+m_{i}^{2}\right) \phi =0  \label{eq:field equations}
\end{equation}
for all $i$ and the total stress-energy-momentum tensor in terms of $\phi $
and $\phi ^{\ast }$ is 
\begin{equation}
T_{\nu }^{\mu }=\sum_{i}\frac{1}{2}\left[ \phi _{,\mu _{i}}^{\ast }\phi
_{,\nu _{i}}-g_{\mu \nu }\left( \phi ^{,\sigma _{i}}\phi _{,\sigma
_{i}}-m_{i}^{2}\phi ^{\ast }\phi \right) +\phi _{,\mu _{i}}\phi _{,\nu
_{i}}^{\ast }\right]
\end{equation}
Each individual part of $T_{\nu }^{\mu }$ will satisfy a conservation
relation since $\phi $ satisfies the field equations (\ref{eq:field
equations})

One thus has 
\begin{equation}
\left( T_{\nu }^{\mu }\right) _{\left( i\right) }W_{\left( i\right) }^{\nu
}=\lambda _{\left( i\right) }W_{\left( i\right) }^{\mu }
\end{equation}
where $W_{\left( i\right) }^{\mu }$ gives the four-velocity of particle $i$.
When the wave function is entangled the individual energy momentum tensors
will be a function of \textit{all} of the particle coordinates and the $%
W_{\left( i\right) }^{\mu }$ will show non local coupling effects even in
the non-interacting particle case. The above may also be extended to include
an external electromagnetic field. Once the state $%
\phi $ is given the stress-energy-momentum tensor for each particle can be
calculated along with its eigenvalues and eigenvectors. As we showed in \cite
{Dewdney96}, for each particle, one finds a pair of eigenvectors one of
which is time-like and the other space-like. The time-like vector and its
eigenvalue determine the flows of energy and the density respectively. As we
have also shown elsewhere, the individual particle velocities using equation
(\ref{eq:PSdecomposition}) are given by 
\begin{equation}
\left( v^{k}\right) _{i}=\left( \frac{S^{k}\pm e^{\pm \theta }\nabla P}{%
-\left( \frac{\partial S}{\partial t}\pm e^{\pm \theta }\frac{\partial P}{%
\partial t}\right) }\right) _{i}  \label{eq:vh}
\end{equation}
where 
\begin{equation}
\left( sinh\theta \right) _{i}=\left( \frac{P^{\mu }P_{\mu }-S^{\mu }S_{\mu }%
}{2P^{\mu }S_{\mu }}\right) _{i}
\end{equation}

It would therefore appear that in the non-interacting field case one can
calculate many-particle trajectories. However, for the case of entangled
wave functions (which imply non-local correlation of the individual particle
trajectories) the prescription given thus far is not complete.

In order to calculate specific trajectories some further structure is
required to specify the coordination of the particles in time. Consider the
entangled two-particle wave function, defined on a four-dimensional
configuration-space time 
\begin{equation}
\Psi \left( x_{1}^{\left( 1\right) },x_{0}^{\left( 1\right) };x_{1}^{\left(
2\right) },x_{0}^{\left( 2\right) }\right) =\phi _{a}\left( x_{1}^{\left(
1\right) },x_{0}^{\left( 1\right) }\right) \phi _{b}\left( x_{1}^{\left(
2\right) },x_{0}^{\left( 2\right) }\right) +\phi _{b}\left( x_{1}^{\left(
1\right) },x_{0}^{\left( 1\right) }\right) \phi _{a}\left( x_{1}^{\left(
2\right) },x_{0}^{\left( 2\right) }\right)  \label{eq:4spaceentagled}
\end{equation}
which may be taken to describe the motion of two non-interacting particles
each confined to one dimension $\left( x_{1}=z\right) ,$ and we shall write $%
x_{0}=t$. In order to begin to calculate the world lines of a particular
configuration of particles four coordinates defining a point in the
configuration space-time must be specified. In the relativistic case, with
two time coordinates, $\left( t^{\left( 1\right) },t^{\left( 2\right)
}\right) $, the combination of this initial point and the wave function (\ref
{eq:4spaceentagled}) is not yet sufficient to define a unique trajectory. One
needs also to specify an extra relationship between $\left( x^{\left(
1\right) },t^{\left( 1\right) }\right) $ and $\left( x^{\left( 2\right)
},t^{\left( 2\right) }\right) .$ That is, using our prescription, although $%
\frac{\partial x_{1}^{\left( 1\right) }}{\partial t^{\left( 1\right) }}$ and 
$\frac{\partial x_{1}^{\left( 2\right) }}{\partial t^{\left( 2\right) }}$
can be calculated, to update the particle positions one then needs to
calculate $\frac{\partial x_{1}^{\left( 1\right) }}{\partial t^{\left(
1\right) }}\triangle t^{\left( 1\right) }$ and $\frac{\partial x_{1}^{\left(
2\right) }}{\partial t^{\left( 2\right) }}\triangle t^{\left( 2\right) }$
and the problem is how to choose the relative sizes of $\triangle t^{\left(
1\right) }$ and $\triangle t^{\left( 2\right) }$ given that different
choices will, with wave functions like (\ref{eq:4spaceentagled}), yield
different trajectories. One needs a covariant prescription.

\section{Lorentz invariance}

The light-cone is a universal invariant local structure which may be
employed to give the desired covariant theory. The invariance of the light
cone can be characterized by the eigenvalues and eigenvectors of the Lorentz
transformation, for simplicity we discuss the case of one spatial dimension.
Writing 
\begin{equation}
\left[ 
\begin{array}{cc}
\cosh \alpha & -\sinh \alpha \\ 
-\sinh \alpha & \cosh \alpha
\end{array}
\right] \left[ 
\begin{array}{c}
a \\ 
b
\end{array}
\right] =\lambda \left[ 
\begin{array}{c}
a \\ 
b
\end{array}
\right]
\end{equation}
we find 
\begin{eqnarray}
a\cosh \alpha -b\sinh \alpha &=&\lambda a \\
-a\sinh \alpha +b\cosh \alpha &=&\lambda b
\end{eqnarray}
hence 
\begin{equation}
\frac{a}{b}=\frac{\sinh \alpha }{\cosh \alpha -\lambda }=\frac{\cosh \alpha
-\lambda }{\sinh \alpha }
\end{equation}
and 
\begin{equation}
\lambda =\cosh \alpha \mp \sinh \alpha =e^{\mp \alpha }
\end{equation}
for $\frac{a}{b}=1$ we have $\lambda =e^{-\alpha }$ and for $\frac{a}{b}=-1$
we have $\lambda =e^{+\alpha }.$ Eigen vectors are null vectors which are
scaled by $e^{\mp \alpha }$ after Lorentz transformation. For relative
velocity $v,$ $\alpha $ is defined by 
\begin{equation}
v=\tanh \alpha
\end{equation}
and we have 
\begin{equation}
\frac{1+v}{1-v}=\frac{\cosh \alpha +\sinh \alpha }{\cosh \alpha -\sinh
\alpha }=e^{2\alpha }
\end{equation}
and 
\begin{eqnarray}
e^{+\alpha } &=&\sqrt{\frac{1+v}{1-v}} \\
e^{-\alpha } &=&\sqrt{\frac{1-v}{1+v}}
\end{eqnarray}
see \cite{Barut}. This can easily be generalized to any homogeneous Lorentz
transformation.

We now proceed as follows. At the location of each particle construct the
local null surfaces (in Minkowski space-time this will be extended over
space-time with coordinates defined as $l=t+z,$ $n=t-z,$ $m=x-iy,$ $%
\overline{m}=x+iy).$ A step of $\varepsilon $ in the rest frame of the
particle becomes 
\begin{eqnarray}
\triangle v &=&\varepsilon \sqrt{\frac{1+v}{1-v}} \\
\triangle u &=&\varepsilon \sqrt{\frac{1-v}{1+v}}
\end{eqnarray}
Since $\triangle u\triangle v=\triangle t^{2}-\triangle z^{2}=\varepsilon
^{2},$ $\varepsilon $ is the proper-time step along the particle
trajectories. At the position of particle $1(2)$ moving with velocity $%
v_{1\left( 2\right) }$ with respect to a given inertial frame:

\begin{itemize}
\item  Advance the position of particle $1$ by 
\begin{eqnarray}
\triangle v_{1} &=&\varepsilon \sqrt{\frac{1+v_{1}}{1-v_{1}}} \\
\triangle u_{1} &=&\varepsilon \sqrt{\frac{1-v_{1}}{1+v_{1}}}
\end{eqnarray}
and advance the position of particle $2$ by 
\begin{eqnarray}
\triangle v_{2} &=&\varepsilon \sqrt{\frac{1+v_{2}}{1-v_{2}}} \\
\triangle u_{2} &=&\varepsilon \sqrt{\frac{1-v_{2}}{1+v_{2}}}
\end{eqnarray}

\item  Recalculate the velocities at $u_{1}+\triangle u_{1,}$ etc. and
repeat the process.
\end{itemize}

If one changes the inertial frame with relative velocity $u=\tanh \beta $
then 
\begin{equation}
e^{\mp \alpha _{1,2}}\rightarrow e^{\mp \alpha _{1,2}+\beta }
\end{equation}
which corresponds to the relativistic formula for the addition of velocities 
\begin{equation}
\frac{v_{1,2}+u}{1+v_{1,2}u}=\frac{\tanh \alpha _{1,2}+\tanh \beta }{1+\tanh
\alpha _{1,2}\tanh \beta }=\tanh \left( \alpha _{1,2}+\beta \right)
\end{equation}
In whatever inertial frame one chooses one considers each one of the
particles at rest with the appropriate $\triangle u,\triangle v$. The
formulation is the same for all inertial frames and hence is covariant. One
has taken equal proper time steps for each particle. In employing this
method we note that for calculational purposes it is possible to stick with
the frame coordinates since 
\begin{eqnarray}
\frac{\triangle u+\triangle v}{2} &=&\triangle t=\varepsilon \cosh \alpha \\
\frac{-\triangle u+\triangle v}{2} &=&\triangle x=\varepsilon \sinh \alpha
\end{eqnarray}

Quantum mechanics provides a given state specified on the relativistic
configuration-space-time (of $4n$ -dimensions) for the many particle system.
In order to provide flow-lines in the relativistic configuration-space-time
we need both the eigen vectors of the stress-energy-momentum tensor and a
rule for coordinating the space-time points on the trajectories of the
particles. In preparing a given state both the space and the time
coordinates of the particles are hidden-variables in the sense that, unlike
in clasical mechanics, we cannot control their values.

\section{Illustrative simple model: bound and separated, entangled particles}

In order to illustrate our approach we shall consider a system consisting of
two massive spin zero bosons of equal mass, moving in one dimension, with
coordinates $\left( z^{\left( 1\right) },t^{\left( 1\right) }\right) $ and $%
\left( z^{\left( 2\right) },t^{\left( 2\right) }\right) $ and described by
the two-particle, multi-time Klein-Gordon equation

\begin{equation}
\left( \square _{1}+\square _{2}\right) \Psi \left( z^{\left( 1\right)
},t^{\left( 1\right) },z^{\left( 2\right) },t^{\left( 2\right) }\right)
=m^{2}\Psi \left( z^{\left( 1\right) },t^{\left( 1\right) },z^{\left(
2\right) },t^{\left( 2\right) }\right)
\end{equation}
where we have taken $\hbar =c=1.$

In our simple model each particle is bound in a separate one-dimensional
potential well, the one dimension is taken to be the $z$ axis, but each well
is located at widely separated positions on the $y$ axis. Apart from
enabling the particles to be described as widely separated in space the $y$
coordinate plays no further role and will not be explicitly referenced in
the wave functions. The two confining potentials are taken to be stationary
in the $\sum $ frame in which we specify the wave function as the entangled
state 
\begin{equation}
\Psi \left( z^{\left( 1\right) },t^{\left( 1\right) },z^{\left( 2\right)
},t^{\left( 2\right) }\right) =\phi _{g}\left( z^{\left( 1\right)
},t^{\left( 1\right) }\right) \phi _{e}\left( z^{\left( 2\right) },t^{\left(
2\right) }\right) +\phi _{e}\left( z^{\left( 1\right) },t^{\left( 1\right)
}\right) \phi _{g}\left( z^{\left( 2\right) },t^{\left( 2\right) }\right)
\end{equation}
where 
\begin{equation}
\phi _{g}\left( z,t\right) =\sqrt{\frac{2}{L}}\sin \left( \frac{\pi z}{L}%
\right) e^{i\omega _{g}t}
\end{equation}
the ground state of the well and 
\begin{equation}
\phi _{e}\left( z,t\right) =\sqrt{\frac{2}{L}}\sin \left( \frac{2\pi z}{L}%
\right) e^{i\omega _{e}t}
\end{equation}
the first excited state. The frequencies are given by 
\begin{eqnarray}
\omega _{g} &=&\sqrt{\left( \frac{\pi }{L}\right) ^{2}+m^{2}} \\
\omega _{e} &=&\sqrt{\left( \frac{2\pi }{L}\right) ^{2}+m^{2}}
\end{eqnarray}
In our approach the individual particle velocities can be calculated
according to equation (\ref{eq:vh}). The velocity of each particle depends on
all of the coordinates $\left( z^{\left( 1\right) },t^{\left( 1\right)
},z^{\left( 2\right) },t^{\left( 2\right) }\right) ,$ but, as discussed
above, the specification of a particular set of values of these coordinates,
along with a velocity formula, although necessary, is not sufficient, to
produce a unique trajectory in the four-dimensional space-time. Our
covariant formulation provides the necessary extra condition to produce a
unique Lorentz-invariant particle trajectory for the system in the
four-dimensional space-time. This trajectory can be projected into a pair of
particle trajectories in any inertial frame. Changing inertial frame simply
amounts to a passive re-coordinatisation of the configuration space-time and
the invariant trajectory. The points $(z^{\left( 1\right) _{j}},t^{\left(
1\right) _{j}})$ on the trajectory of particle $1$ are correlated with the
points $(z^{\left( 2\right) _{j}},t^{\left( 2\right) _{j}})$ on the
trajectory of particle $2$ in a unique and Lorentz invariant manner. That is
to say that, if the calculation is carried out using a different inertial
frame, $\sum^{\prime }$ say, then the transformed wave function and
transformed initial coordinates produce trajectories which are the Lorentz
transforms of those in $\sum $ and the individual points on the particle
trajectories are correlated in the same way. There is now a unique velocity
at each point in the four-dimensional configuration space-time of the
two-particle system, hence there is only one trajectory passing through each
point.

Let us consider a specific case, if $t^{\left( 1\right) _{0}}=t^{\left(
2\right) _{0}}=0,$ the particles start from equal times (in the $\sum $
frame), the three-velocities of the particles in $\sum $ are both zero and
the particles remain coordinated at equal times in this frame. This is shown
in figure 1, where the initial point is taken as $(1.0,0.0,2.0,0.0)$ (in the 
$\sum $ frame) and the coordination of the points on the trajectories of
each particle is indicated by the numbering. In figure 2 the initial point
was taken as $(1.0,1.0,2.0,0.0)$ (in the $\sum $ frame), the
three-velocities are no longer zero and the points do not remain coordinated
at a constant time-difference.

Covariance is assured since we use the proper times of the particles to
establish the coordination of the points on the individual particle
trajectories. One may choose an ensemble of trajectories that cross the $%
t^{\left( 1\right) }=t^{\left( 2\right) }=0$ hyperplane and this would
correspond with the set of trajectories which are at equal times, namely $%
t=0 $, in $\sum .$ One is not bound to make such a choice of initial points
but once such an ensemble is agreed upon all observers would calculate the
same trajectories. This means that, given the hidden variables, the outcomes
of any measurements are completely determined, irrespective of the frame of
reference in which the motion is described. In the context of this simple
model one can imagine measuring the energy of the particles by removing the
confining potentials and allowing the particles to run out. A time-of-flight
measurement then reveals the energy of the particle. Consider the case in
which the potentials are removed at equal times $t^{\left( 1\right)
}=t^{\left( 2\right) }=\tau $ in $\sum .$ If the particles cross the $%
t^{\left( 1\right) }=t^{\left( 2\right) }=0$ hyperplane then they reach the $%
t^{\left( 1\right) }=t^{\left( 2\right) }=\tau $ hyperplane at equal proper
times. The trajectories, and hence the measurement outcomes at this time,
can be predicted from the fact that configuration-space trajectories cannot cross, we find that if $\left|
z^{\left( 2\right) }\right| >\left| z^{\left( 1\right) }\right| $ then
particle two moves away quickly and hence is found to be the excited
particle whilst particle one moves slowly and is found to be the ground
state particle. On the other hand if $\left| z^{\left( 2\right) }\right|
<\left| z^{\left( 1\right) }\right| $ then the outcome is reversed; particle
one moves away quickly and hence is found to be the excited particle whilst
particle two moves slowly and is found to be the ground state particle.
Given the initial point in configuration space-time, a unique outcome is
guaranteed. As we have emphasised above adopting a different frame amounts
to a passive re-coordinatization of the space-time. Different outcomes would
only be predicted if the initial point in the configuration space-time was
taken to be different and there is no reason to change this point just
because a different frame for the description of the experiment has been
adopted.

\section{Conclusion}

We have shown that there is no necessary contradiction between relativistic
quantum mechanics and a Lorentz-invariant hidden-variable theory based on
particle trajectories. The theory that we propose requires one extra Lorentz
invariant rule which produces unique flow-lines in the many-particle
relativistic configuration space. The extra rule applies equally to bosons
and fermions. As is usual in the de Broglie-Bohm approach to quantum
mechanics, measurement plays no special role; rather the introduction of a
measuring device merely enlarges the configuration space in which the whole
system is described.

\bigskip

\newpage {\bf Figure Captions}

\begin{enumerate}
\item  FIG. 1. The initial point is taken as $(1.0,0.0,2.0,0.0)$ (in the $%
\sum $ frame) and the coordination of the points on the trajectories of each
particle is indicated by the numbering.

\item  FIG. 2. The initial point was taken as $(1.0,1.0,2.0,0.0)$ (in the $%
\sum $ frame), the three-velocities are no longer zero and the particles do
not remain coordinated at a constant time-difference.
\end{enumerate}

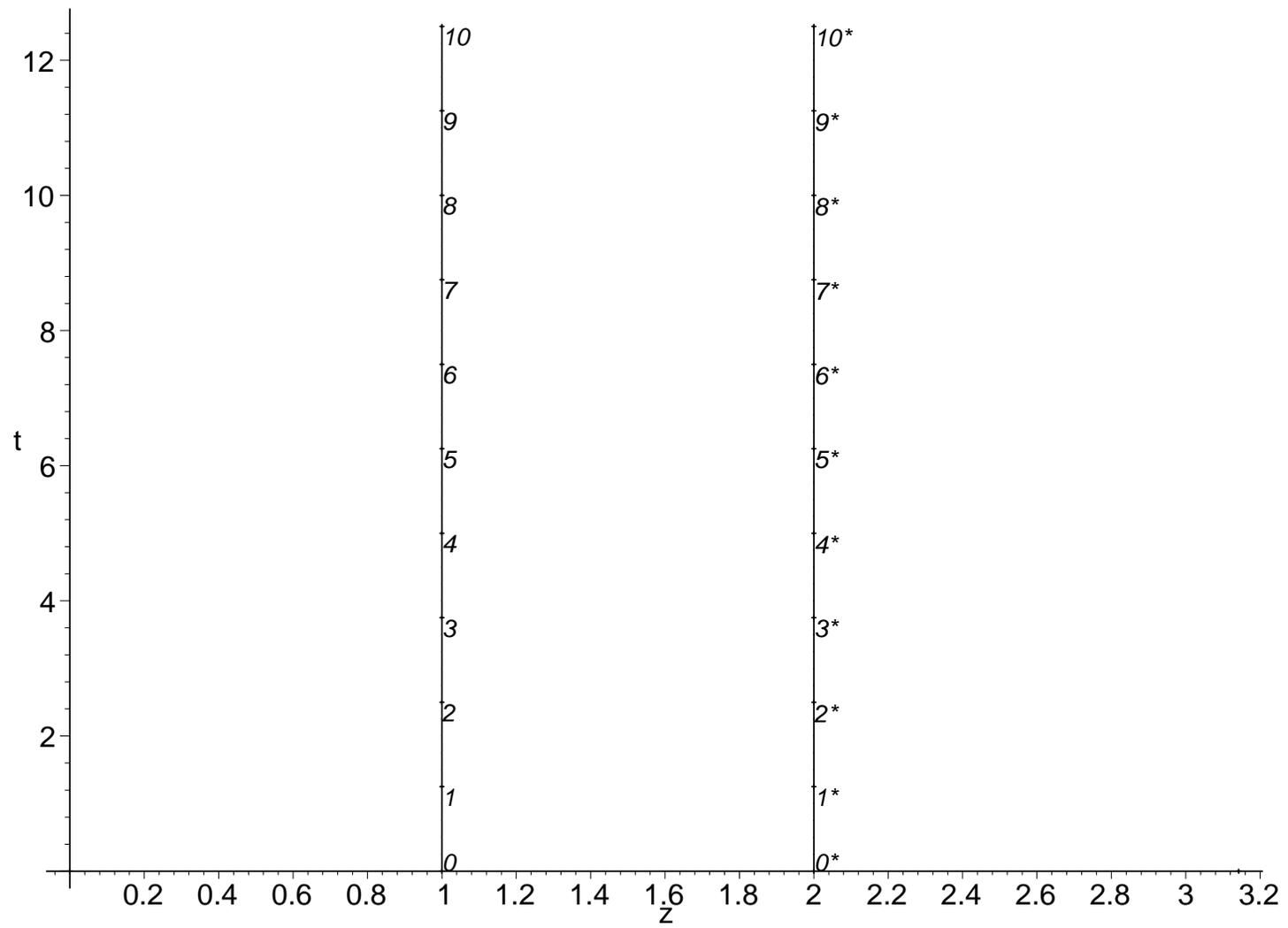
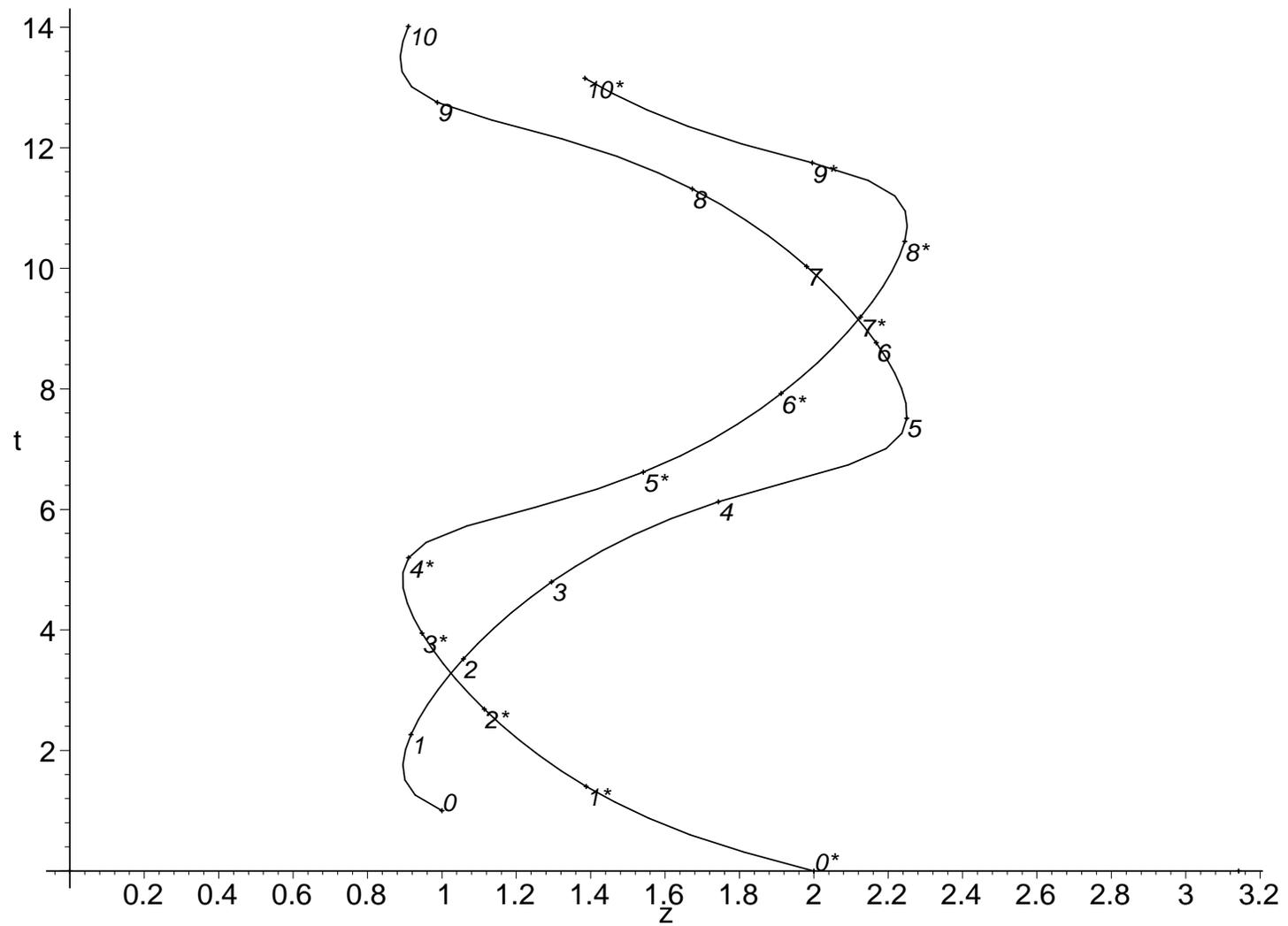
\end{document}